# Distributed NLP

*Galip Aydin and Ibrahim Riza Hallac
Faculty of Engineering, Department of Computer Engineering Firat University, Turkey

**Abstract**

In this paper we present the performance of parallel text processing with Map Reduce on a cloud platform. Scientific papers in Turkish language are processed using Zemberek NLP library. Experiments were run on a Hadoop cluster and compared with the single machine's performance.

## 1. Introduction

In this study a text processing work using a Natural Language Processing library is done for automatically simplifying scientific papers written in Turkish language. The aim is to remove non-informative words in the documents so that unnecessary parts, which can be called noise, are eliminated. Traditional approaches for this operation are not very useful when there is hundreds of thousands or millions of input files. Therefore we implemented a parallel processing solution to this problem. In this work, text processing is done in parallel using Map Reduce programming model. Hadoop framework on top of a cloud environment is used for the test bed. Two applications were run for the comparison of the parallel and standard techniques performance. Parallel application was compared in itself with file split sizes. A natural language processing library for Turkish language called Zemberek is used to determine which word is the adjective, adverb, prepositional.

## 2. Background

### 2.1. MapReduce

In 2004, Google published the MapReduce paper[1] which demonstrated a new type of distributed programming model that makes it easy to run high performance parallel programs on big data using commodity hardware.

Basically MapReduce programs consists of two major modules; mappers and reducers. A typical MapReduce job works as follows: First data is split into blocks then is sent to the mappers. Mappers do their part by creating (key, value) pairs from the input. The framework then generates key,value pair lists as the number of keys output from mappers.

*Corresponding author: Address: Faculty of Engineering, Department of Computer Engineering Firat University, 23100, Elazig TURKEY. E-mail address: gaydin@firat.edu.tr, Phone: +904242370000 /6313

Map (k1,v1) → list (k2,v2).

After these so called intermediate key/value lists are produced, they are sent to the reducers. A reducer is responsible from creating the sum of the intermediate keys/values, which can be seen as a grouping operation:

Reduce (k2, list (v2)) → list (v3).

Mappers and reducers are user defined programs which are implemented by using the MapReduce API. Therefore a MapReduce job is composed of several processes such as splitting and distributing the data, map and reduce codes and writing results to the distributed file system etc. Sometimes analyzing data using MapReduce may require running more than one job. The jobs can be independent from each other or they may be chained for more complex scenarios.

## 2.2. Apache Hadoop

In this work we used the most popular open source implementation of Google's MapReduce framework called  Hadoop. Hadoop enables storing and processing big data using the MapReduce programming model [2].

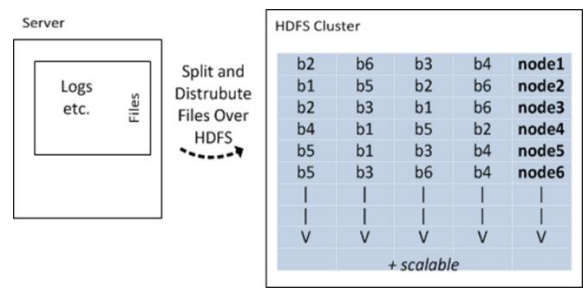

**Figure 1.** HDFS

Hadoop uses master –slave architecture. Name Node and Job Tracker are master nodes whereas Data Node and Task Tracker are slave nodes of the cluster [3]. The input data is partitioned into blocks and this blocks are placed into Data Nodes. Name Node holds the metadata of the blocks so the Hadoop system knows which block is stored on which Data Node and if one node fails it doesn't spoil the completion of the job because Hadoop knows where the replicas of the those blocks are stored.

Job Tracker and Task Tracker tracks the execution of processes. They have the similar relation with Name Node and Data Node. Task tracker is responsible for running the tasks and sending messages to Job Tracker. Job Tracker communicates with Task Tracker and keeps record of running processes. If job tracker detects that a task tracker is failed or unable to complete its part of the job, it schedules the missing executions on another task tracker [4].

*2.3. Hadoop HDFS*

Hadoop uses The Hadoop Distributed File System (HDFS) which is the open source version of Google File System [5] . The data in HDFS is stored on a block-by-block basis. First the files are split into blocks, then are distributed over the Hadoop Cluster. Each block in the HDFS is 64 MB by default unless the block size has been modified by the user. The default block size can be changed from the configuration file with changing the value of dfs.block.size parameter. If the file is larger than 64 MB the HDFS splits it from a line where the file size doesn't exceed the maximum block size and the rest of the lines (for text input) in the file are moved to a new block. Figure-2 shows that the input data is split into the blocks (b1,b2,b3 etc) and distributed over the nodes (node1,node2 etc) in the cluster. As shown in the Figure-1 a block is not stored only in one specific node. This is because Hadoop stores replicas of the blocks in three different nodes by default. The replication factor can be changed by configuring the dfs.replication value in hdfs-default.xml file.

*2.4. Cloud Computing*

"Like a traditional Operating System (OS), a cloud OS is responsible for managing the low level cloud resources and presenting a high level interface to the application programmers in order to hide the infrastructure details. However, unlike a traditional OS, a cloud OS has to manage these resources at scale. It is far from obvious that we can simplify large-scale systems' design and implementation if we build them on top of a cloud OS. The tradeoffs a cloud has made in favor of"[6]

Hadoop cluster can be set up by installing and configuring necessary files on commodity servers. However it can be a daunting and challenging work when there are hundreds or even thousands of servers to be used as Hadoop nodes in a cluster. Cloud systems provides an infrastructure which is easy to scale, easy to manage network and storage, and fault tolerant features.

Geoffrey Fox and his friends show the advantages and challenges of running MapReduce in cloud environments [6]. "In fact, the utility computing model offered by cloud computing is remarkably well-suited for scientists' staccato computing needs. While clouds offer raw computing power combined with cloud infrastructure services offering storage and other services, there is a need for distributed computing frameworks to harness the power of clouds both easily and effectively. At the same time, it should be noted that cloud infrastructures are known to be less reliable than their traditional cluster counterparts and do not provide the high-speed interconnects needed by frameworks such as MPI."

There are several options for setting up a Hadoop cluster. Paid cloud systems like Amazon EC2 provides EMR clusters for running MapReduce jobs [7]. In EC2 cloud the input data can be distributed to Hadoop nodes through uploading files over the master node. Because pricing in the clouds is on a pay as go basis, customers don't have to pay for the idle nodes. Amazon shuts down the rented instances after the job completes. In this case, all the data will be removed from the system. For example, if the user wants to run another job over the pre-used data he/she has to upload it again. If data is stored on Amazon Simple Storage Service (Amazon S3) user can use it as long as he/she pays for the storage [8]. Amazon also provides some facilities for monitoring working Hadoop jobs. There is another option for setting up a Hadoop cluster with manually configuring instances. The instances can be servers connected by a network or they can be instances of a cloud system.

OpenStack is an open-source cloud management platform which can be used as an Infrastructure as a Service software [9]. One can easily set up and manage a cloud system with installing OpenStack on the first layer of their operating system. The OpenStack platform used in the study was provided by FutureGrid project. Future Grid provides experimental computing grid and cloud test-bed to the research community. 7 instances of virtual machines were created to be nodes of Hadoop cluster on OpenStack Grizzly. Their specifications are shown in the diagram below.

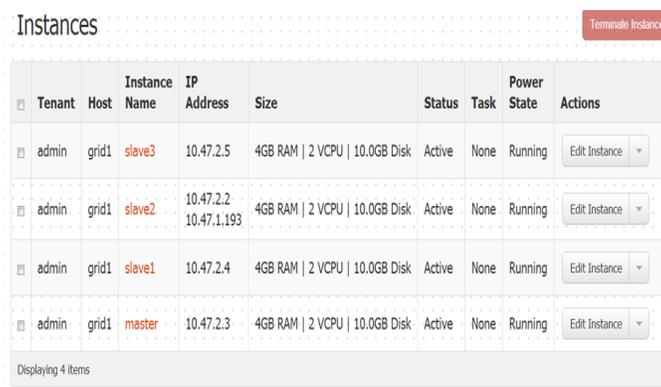

**Figure 2.** OpenStack

## 2.5. Cloud Computing

Bigdata is a horizontally scaled storage and computing fabric supporting optional transactions, very high concurrency, and very high aggregate IO rates. Bigdata was designed from the ground up as a distributed database architecture running over clusters of 100s to 1000s of machines, but can also run in a high-performance single-server mode.
The bigdata architecture provides a high-performance platform for data-intensive distributed computing, indexing, and high-level query on commodity clusters. While the semantic web database layer has received the most attention, the bigdata architecture is well suited for a wide range of data models, workloads, and applications.

## 2.6. Zemberek

In this study we used Zemberek  Natural Language Processing library alongside with Hadoop MapreReduce framework. Zemberek is an open source project written in Java programming language. This toolset is very useful for Turkic languages, especially Turkish [10].

## 3. Architecture and Implementation

When extracting useful information from the text, one should apply approaches like machine learning in a very large scale. Otherwise regardless of how efficient the algorithm is, obtaining information from the text data generally fails if the process is not done on an enough amount of data.

Processing large amounts of data with the traditional programming approaches needs large computational power. Instead of using expensive solutions like a super computer, one can use grid environments. Cloud computing is the most popular emerging grid environment in recent years which is highly used in many big data problems. Cloud computing provides an environment which runs applications in a scalable, utilizable basis. Applications in this study were run on a cloud environment which is called OpenStack.

For parallel processing, we used Apache's Map Reduce implementation, Hadoop. Hadoop was installed on virtual machines which were deployed on Open Stack Cloud Platform. The Open Stack Cloud was provided by Future Grid [11] Sierra Platform. The structure of Hadoop platform created for this study can be shown as in Figure 4. On each node Ubuntu 13.04 is installed. On each machine Hadoop 1.0.3 Release with JDK 7 were set up.

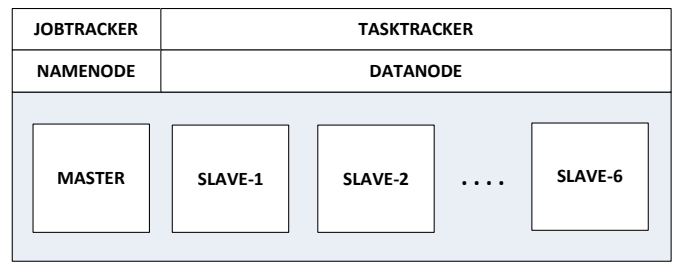

**Figure 3.** Hadoop cluster

The architecture of the complete system is shown in Figure 5. There is the FutureGrid OpenStack platform on the bottom where the virtual machines are running on. Parallel text processing applications uses data stored in HDFS as input. Console applications uses the data stored in the used operating system's (Ubuntu) default file system. The Zemberek Natural Language Processing libraries are called both in the MapReduce and in the console applications.

In this work we implemented different applications for comparing the performance of text processing between standard programming and MapReduce model. Different test cases formed for comparing the performance of parallel text processing jobs over standard computing. Input files were created from academic papers written in Turkish Language. Papers were downloaded in pdf format and converted to text files using PDF-Box Java Library. The input files were arranged to be in 100MB, 1GB, 2GB, 4GB, 8GB and 16GB sizes. Jobs run on these different sizes of inputs both on "bare metal, virtualized" single computer and 7 machines in parallel. All machines had the same capacity for memory and processor. (4GB RAM | 2 VCPU | 40GB Disk).

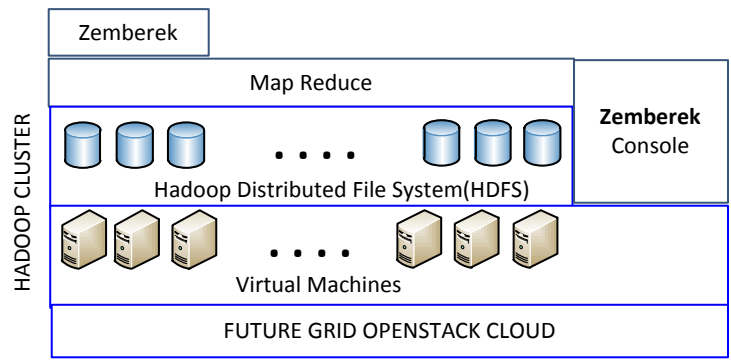

**Figure 4.** Architecture

In the map step of the MapReduce applications the lines are processed with the Zemberek libray. A sample input/output for Zemberek is shown in Table 1. The MapReduce applications in this test can be called as Map-Only Jobs. Mapped data is written to output directory without performing the reduce step. Flowcharts of the MapReduce and Console Applications are shown in Figure 5 and Figure 6.

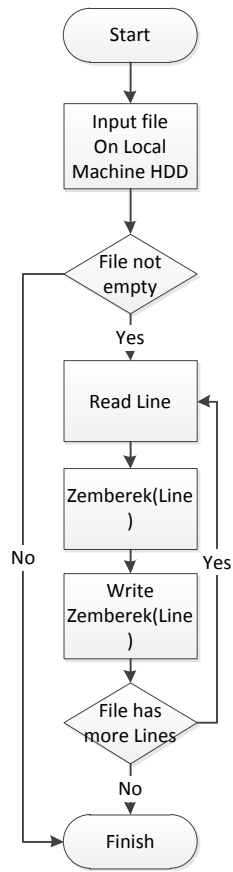 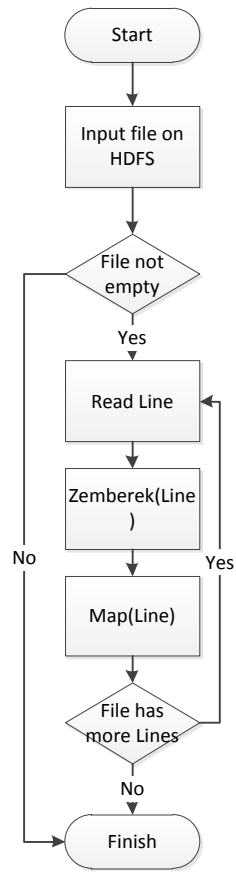

**Figure 5.** Console application    **Figure 6.** Hadoop application

**Table 1.** Sample text

| |
|---|
| Feshe itiraz davası, işverence geçerli sebep gösterilmeden ya da kanunda öngörülen usule uyulmadan yapılan fesihlere karşı işçilerin başvurabileceği bir itiraz yolu olarak karşımıza çıkmaktadır. |
| Feshe itiraz dava işveren geçerli sebep gösterilmeden kanun öngör usul uyulmadan fesih işçi başvur itiraz yol olarak karşı |

## 4. Results

Performance comparison of the applications were done based on their running times. In order to obtain accurate running times, each test was performed at least ten times. Longest and shortest running times were omitted from these ten running times. Average of the rest of the results determined as "running time".

Hadoop Job implemented as Map Only Jobs. Lines from files were processed and were mapped and without the reduce step they were written to output files. Console Job implemented using standard Java File Read/Write methods.

Single file tests are which the input data used in the application is a complete file. In Hadoop applications, this files were distributed to the nodes with the replication factor of three. In console applications this files are directly read by the operating system. Multi file tests are which the input data pieced to 100 MBs of files. By this way, performances of applications are tested according to structure of their input data.

**Table 2.** Running times of the experiments

| | Running Time (minutes) | |
|---|---|---|
| Size (GB) | Console Test (single machine) | Hadoop Test (a cluster of 7 machines) |
| 0,1 | 2,3 | 1,33 |
| 1 | 23,3 | 3,35 |
| 2 | 46,7 | 5,26 |
| 4 | 92,7 | 9,1 |
| 8 | 183,3 | 17,16 |
| 16 | 383,9 | 33,3 |

Running times of the console tests are showed in both as table and graphics in Table 2, Figure 7 and Figure 8. It can be seen from the results of the running times that running this text processing program on a single machine is not convenient when there is large amount of data. Performance of the parallel application increases when the size of input increases. In this study, performance of Hadoop Jobs are also compared based on the split sizes of the input data. First the Jobs are run with an input of a single file, than the Jobs are run with the same size of input file but in 100 MB splits. It's seen that performance of the Jobs is inversely proportional to the input split number.

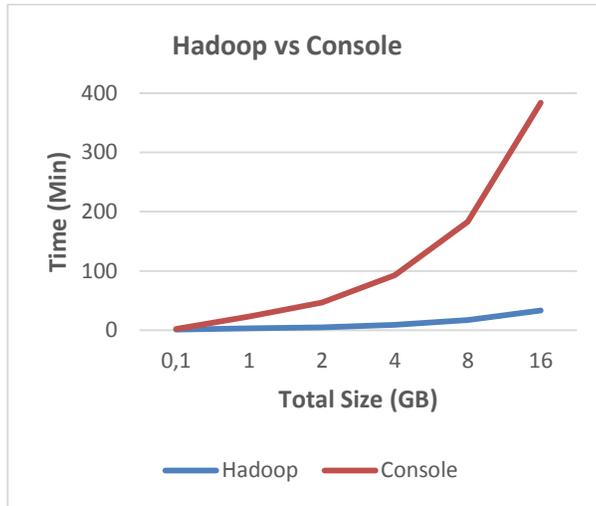

**Figure 7.** Hadoop vs console test

**Table 3.** Single file vs multi file test values

| Size | Time (sec) | |
| --- | --- | --- |
| (GB) | Single file | Multi file |
| 0,1 | 1,33 | 1,1 |
| 1 | 3,35 | 1,29 |
| 2 | 5,26 | 5,17 |
| 4 | 9,1 | 9,52 |
| 8 | 17,16 | 18,5 |
| 16 | 33,3 | 36,045 |

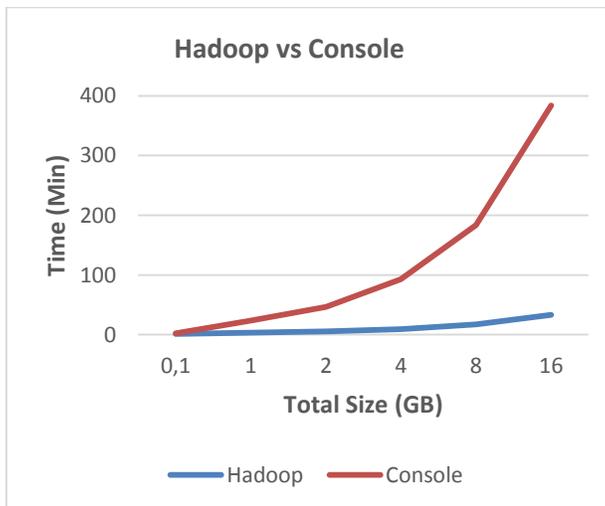

**Figure 8.** Single file vs multi file test

It is stated in HIPI [12] that the performance of Hadoop is significantly better when the input is a large file instead of many small files. Distributing files over the nodes takes time and when the number of files increases it can take about a hundred times longer comparing to distributing a big large file.

**Conclusions**

Over the last few years, cloud computing technologies have been widely used in the process of many different big data problems. MapReduce is now considered as a very efficient programming model for processing files in parallel. By doing this study we examined MapReduce with using external libraries. All the tests in this study were run on an open source cloud-computing platform.

-